\documentclass{PoS}
\PoS{PoS(LAT2005)184}
\title{$J/\Psi$ at high temperatures in anisotropic lattice QCD}

\ShortTitle{$J/\Psi$ at high temperatures in anisotropic lattice QCD}

\author{\speaker{Hideaki Iida} and Noriyoshi Ishii\\

        Department of Physics, H-27, Tokyo Institute of Technology,\\
Oh-okayama 2-12-1, Meguro, Tokyo 152-8551, Japan\\
         E-mail: \email{iida@th.phys.titech.ac.jp}, 
                 \email{ishii@th.phys.titech.ac.jp}}

\author{ Takumi Doi 

        \\

 RIKEN BNL Research Center, Brookhaven National Laboratory, \\
Upton, New York 11973, USA\\
         E-mail: \email{doi@quark.phys.bnl.gov}}

\author{ Hideo Suganuma

        \\

 Department of Physics, Kyoto University, \\
Kitashirakawaoiwake, Sakyo, Kyoto 606-8502, Japan\\
         E-mail: \email{suganuma@scphys.kyoto-u.ac.jp}}

\abstract{
$J/\Psi$ and $\eta_c$ above the QCD critical temperature $T_c$ 
are studied in anisotropic quenched lattice QCD, 
considering whether the $c\bar c$ systems above 
$T_c$ are compact quasi-bound states or scattering states. 
We adopt the standard Wilson gauge action and $O(a)$-improved Wilson 
quark action with renormalized anisotropy $a_s/a_t =4$ 
at $\beta=6.10$ on $16^3\times (14-26)$ lattices, 
which correspond to the spatial lattice volume 
$V\equiv L^3\simeq(1.55{\rm fm})^3$ and temperatures $T\simeq(1.11-2.07)T_c$. 
To clarify whether compact charmonia survive in the deconfinement phase, 
we investigate spatial boundary-condition dependence of the energy of the 
$c\bar c$ systems above $T_c$.
In fact, for low-lying $c \bar c$ scattering states, 
there appears a significant energy difference $\Delta E \equiv E{\rm (APBC)}-E{\rm (PBC)}$ 
between periodic and anti-periodic boundary conditions 
as $\Delta E\simeq2\sqrt{m_c^2+3\pi^2/L^2}-2m_c$ ($m_c$: charm quark mass) 
on the finite-volume lattice.
In contrast, for compact charmonia, there is no significant energy difference 
between periodic and anti-periodic boundary conditions. 
As a lattice QCD result, we find 
almost no spatial boundary-condition dependence for the energy of the $c\bar c$ system 
in $J/\Psi$ and $\eta_c$ channels  
for $T\simeq(1.11-2.07)T_c$, which indicates that 
$J/\Psi$ and $\eta_c$ would survive as compact $c\bar c$ quasi-bound states below $2T_c$. 
}
\FullConference{XXIIIrd International Symposium on Lattice Field Theory\\

		 25-30 July 2005\\

		 Trinity College, Dublin, Ireland}

\begin{document}
\section{Introduction}
Since QCD is established, the quark-gluon-plasma (QGP) phase 
has been studied with much attention as a ``new phase of matter" at high temperatures  
both in theoretical and experimental sides \cite{HMHK86,MS86,NA50,RHIC}. 
In recent years, QGP creation experiments are actually performed at 
SPS \cite{NA50} and RHIC \cite{RHIC} in high-energy heavy-ion collisions.
As an important signal of the QGP creation, 
$J/\Psi$ suppression \cite{HMHK86,MS86} was theoretically proposed 
and has been tested in the SPS/RHIC experiments. 
The basic assumption of $J/\Psi$ suppression is that 
$J/\Psi$ disappears above $T_c$ due to vanishing of 
the confinement potential and the Debye screening effect \cite{MS86}. 

Very recently, some lattice QCD calculations indicate an interesting possibility that 
$J/\Psi$ and $\eta_c$ seem to survive even above $T_c$ \cite{UKMM01,AH04,DKPW04},
which may lead a serious modification for the $J/\Psi$ suppression scenario 
in QGP physics.
However, as a possible problem, the observed $c\bar c$ state on lattices 
may not be a nontrivial charmonium but a trivial $c\bar c$ scattering state, 
because it is difficult to distinguish these two states in lattice QCD.

In this paper, we aim to clarify whether the $c\bar c$ system above $T_c$ 
is a compact quasi-bound state or a scattering state, which is spatially spread. 
To distinguish these two states, we investigate spatial boundary-condition dependence
of the energy of the $c\bar c$ system by 
comparing results in periodic and anti-periodic boundary conditions.
If the $c\bar c$ system is a scattering state,
there appears an energy difference $\Delta E$ between the 
 two boundary conditions as  
$\Delta E\simeq 2\sqrt{m_c^2+3\pi^2/L^2}-2m_c$
 with the charm quark mass $m_c$ on a finite-volume lattice with $L^3$. 
 If the $c\bar c$ system is a compact quasi-bound state,
 the boundary-condition dependence is small even in finite volume. 
 In Ref.\cite{IDIOOS05}, this method is actually applied for distinction between 
a scattering state and a compact resonance.

\section{Method to distinguish a compact state from a scattering state}
To begin with, we briefly explain the method to distinguish a compact state from 
a scattering state 
in term of its spatial extension.
For this purpose, we investigate the $c \bar c$ system 
in the periodic boundary condition (PBC)  
and in the anti-periodic boundary condition (APBC), respectively, 
and examine spatial boundary-condition dependence for the $c \bar c$ system. 
Here, in the PBC/APBC case, we impose periodic/anti-periodic boundary condition for $c$ and $\bar c$ 
on a finite-volume lattice.

\begin{table}[hb]
\begin{center}
\label{boundary}
\caption{
Periodic boundary condition (PBC) and anti-periodic boundary condition (APBC): 
the relation between spatial boundary condition and the minimum momentum 
$|\vec p_{\rm min}|$ of $c$, $\bar c$ 
and compact charmonia $c\bar c$.
}
\vspace{0.2cm}

\begin{minipage}{7cm}
\begin{tabular}{ccc}
\hline
\hline
\multicolumn{3}{c}{\bf PBC}\\
\hline
 particle&spatial BC & $|\vec p_{\rm min}|$\\
\hline
$c$ & periodic & $0$ \\
$\bar c$& periodic & $0$\\
\hline
charmonia ($c\bar c$) & periodic & $0 $\\
\hline
\hline
\end{tabular}
\end{minipage}
\begin{minipage}{7cm}
\label{table1} 
\begin{tabular}{ccc}
\hline
\hline
\multicolumn{3}{c}{\bf APBC}\\
\hline
 particle& spatial BC & $|\vec p_{\rm min}|$\\
\hline
$c$ & anti-periodic & $\sqrt{3}\pi /L$ \\
$\bar c$ & anti-periodic & $\sqrt{3}\pi /L$\\
\hline
charmonia ($c\bar c$)& periodic & $0 $\\
\hline
\hline
\end{tabular}
\end{minipage}
\end{center}
\end{table}

\vspace{-0.4cm}

For a compact $c\bar c$ quasi-bound state, the wave function of each quark is spatially localized 
and insensitive to spatial boundary conditions in lattice QCD, so that 
the charmonium behaves as a compact boson 
and its energy in APBC is almost the same as that in PBC \cite{IDIOOS05}.  
For a $c\bar c$ scattering state, 
both $c$ and $\bar c$ have non-zero relative momentum 
$\vec p_{\min}=(\pm\frac{\pi}{L},\pm\frac{\pi}{L},\pm\frac{\pi}{L})$, i.e.,
$|\vec p_{\rm min}|=\sqrt{3}\pi/L$ in APBC, 
while they can take zero relative momentum $\vec p_{\rm min}=0$ in PBC.
In fact, if the $c\bar c$ system is a scattering state, 
there appears a significant energy difference 
$\Delta E$ between PBC and APBC due to the finite lattice volume of $L^3$,
and it is estimated as $\Delta E \simeq 2\sqrt{m_c^2+3\pi^2/L^2}-2m_c$. 
In our lattice QCD calculation, 
$|\vec p_{\rm min}|$ and $\Delta E$ for the $c \bar c$ scattering state are estimated as 
$|\vec p_{\rm min}| =\sqrt{3} \pi/L \simeq 0.69{\rm GeV}$ and 
$\Delta E \simeq 2\sqrt{m_c^2+3\pi^2/L^2}-2m_c\simeq 0.35{\rm GeV}$ for 
$L \simeq 1.55 {\rm fm}$ and $m_c \simeq 1.3{\rm GeV}$.

\section{Anisotropic lattice QCD}
In this paper, we adopt anisotropic lattice QCD for the study of high-temperature QCD.
In lattice QCD at temperature $T$, 
(anti)periodicity is imposed in the temporal direction with the period $1/T$, 
and hence it is technically difficult to measure temporal correlators at high temperatures.
To overcome this problem, we use the anisotropic lattice with anisotropy $a_s/a_t=4$. 
Owing to the finer temporal mesh, we can obtain 
detailed information for temporal correlators. 
 
For the gauge field, we adopt the standard plaquette action on an anisotropic lattice 
as \cite{IDIOOS05,MOU01}
\begin{eqnarray}
S_G=\frac{\beta}{N_c}\frac{1}{\gamma_G}
\sum_{s,i<j\leq 3} {\rm Re}{\rm Tr}\{1-P_{ij}(s)\}
+\frac{\beta}{N_c}\gamma_{G}\sum_{s,i\leq 3}{\rm Re}{\rm Tr}\{1-P_{i4}(s)\},
\end{eqnarray}
where $P_{\mu\nu}$ denotes the plaquette operator.
In the simulation, we take $\beta \equiv 2N_c/g^2=6.10$ and  
the bare anisotropy $\gamma_G=3.2103$, 
which lead to renormalized anisotropy as $a_s/a_t=4.0$.
The scale is set by the Sommer scale $r_0^{-1}=395{\rm MeV}$. 
Then, the spatial and temporal lattice spacing are estimated as 
$a_s^{-1}\simeq 2.03{\rm GeV}$ (i.e., $a_s\simeq$ 0.097fm),
and $a_t^{-1}\simeq 8.12{\rm GeV}$ (i.e., $a_t\simeq$ 0.024fm), respectively.
The adopted lattice size is $16^3\times (14-26)$, 
which corresponds to the spatial lattice size as $L \simeq 1.55{\rm fm}$ 
and the temperature as $T=(1.11-2.07)T_c$. 
We use 999 gauge configurations, which are picked up every 
500 sweeps after the thermalization of 20,000 sweeps. 
 
For quarks, we use $O(a)$-improved Wilson (clover) action 
on the anisotropic lattice as \cite{IDIOOS05,MOU01}
\begin{eqnarray}
S_F &\equiv& \sum_{x,y}\bar \psi(x)K(x,y)\psi(y), \nonumber \\
K(x,y) \equiv \delta_{x,y}&-&\kappa_t\{(1-\gamma_4)U_4(x)\delta_{x+\hat 4,y}
+(1+\gamma_4)U_4^\dagger(x-\hat 4)\delta_{x-\hat 4,y}\}\nonumber \\
&-& \kappa_s \sum_i \{ (r-\gamma_i)U_i(x)\delta_{x+\hat i,y}\}
+(r+\gamma_i)U_i^\dagger(x-\hat i)\delta_{x-\hat i,y}\} \nonumber \\
&-&\kappa_s c_E\sum_i \sigma_{i4}F_{i4}\delta_{x,y}-r\kappa_s c_B\sum_{i<j}
\sigma_{ij}F_{ij}\delta_{x,y},
\end{eqnarray}
which is anisotropic version of the Fermilab action \cite{EKM97}.
$\kappa_s$ and $\kappa_t$ denote the spatial and temporal hopping 
parameters, respectively, and $r$ the Wilson parameter. 
$c_E$ and $c_B$ are the clover coefficients.
The tadpole improvement is done by the replacement of $U_i(x)\rightarrow U_i(x)
/u_s$, $U_4(x)\rightarrow U_4(x)/u_t$, where $u_s$ and $u_t$ are 
the mean-field values of the spatial and the temporal link variables, respectively. 
The parameters $\kappa_s, \kappa_t, r, c_E, c_B$ are to be tuned 
so as to keep the Lorentz symmetry up to $O(a^2)$.
At the tadpole-improved tree-level, this requirement leads to 
$r=a_t/a_s$, $c_E=1/(u_su_t^2)$, $c_B=1/u_s^3$ and 
the tuned fermionic anisotropy $\gamma_F\equiv (u_t \kappa_t)/(u_s \kappa_s)=a_s/a_t$.
For the charm quark, we take $\kappa=0.112$ with $1/\kappa \equiv 1/(u_s \kappa_s)-2(\gamma_F+3r-4)$, 
which corresponds to the hopping parameter in the isotropic lattice.
The bare quark mass $m_0$ in spatial lattice unit is expressed 
as $m_0=\frac12(\frac{1}{\kappa}-8)$.
We summarize the lattice parameters and related quantities in Table~2.
In the present lattice QCD, the masses of $J/\Psi$ and $\eta_c$ are found to be  
$m_{J/\Psi} \simeq$ 3.07GeV and $m_{\eta_c} \simeq$ 2.99GeV at zero temperature.

\begin{table}[ht]
\begin{center}
\label{parameters}
\caption{Lattice parameters and related quantities in our anisotropic lattice QCD calculation.}
\vspace{0.2cm}
\begin{tabular}{ccccccccc}
\hline
\hline
$\beta$  & lattice size & $a_s^{-1}$ & $a_t^{-1}$ &$\gamma_G$ & $u_s$ & $u_t$ & $\gamma_F$ & $\kappa$ 
\\
\hline
6.10 & $16^3 \times(14-26)$& 2.03GeV & 8.12GeV&3.2103 &0.8059 &0.9901 &4.0 & 0.112\\
\hline
\hline
\end{tabular}
\end{center}
\end{table}

\section{Temporal correlators of $c\bar c$ systems at finite temperature on
 anisotropic lattice} 

To investigate the low-lying state at high temperatures from the temporal correlator, 
it is practically desired to use a ``good" operator with a large ground-state overlap, 
due to limitation of the temporal lattice size. 
To this end, we use a spatially-extended operator 
of the Gaussian type as
\begin{eqnarray}
O(t,\vec x)\equiv N \sum_{\vec y}\exp\left\{-\frac{|\vec y|^2}{2\rho^2}\right\}
\bar c(t,\vec x+\vec y)\Gamma c(t,\vec x)
\end{eqnarray}
in the Coulomb gauge \cite{IDIOOS05,MOU01}. 
$N$ is a normalization.
The size parameter $\rho$ is optimally chosen in terms of the ground-state overlap. 
$\Gamma=\gamma_k (k=1-3)$ and $\Gamma=\gamma_5$ 
correspond to  $1^- (J/\Psi)$ and $0^- (\eta_c)$ channels, respectively. 
The energy of the low-lying state is calculated from 
the temporal correlator, 
\begin{eqnarray}
G(t)\equiv \frac{1}{V}\sum_{\vec x}\langle O(t,\vec x)
O^\dagger (0,\vec 0)\rangle,
\end{eqnarray}
where the total momentum of the $c\bar c$ system is projected to be zero.

In accordance with the temporal periodicity at finite temperature, 
we define the effective mass $m_{\rm eff}(t)$ 
from the correlator $G(t)$ by the cosh-type function as \cite{ISM02}
\begin{eqnarray}
\frac{G(t)}{G(t+1)}=\frac{\cosh [m_{\rm eff}(t)(t-N_t/2)]} 
{\cosh [m_{\rm eff} (t) (t+1-N_t/2)]}
\label{cosh}
\end{eqnarray}
with the temporal lattice size $N_t$.
In the plateau region of $m_{\rm eff}(t)$, 
$m_{\rm eff}(t)$ corresponds to the energy of the low-lying $c\bar c$ state.
To find the optimal value of $\rho$, 
we calculate the correlator $G(t)$ for $\rho=0.2, 0.3, 0.4$ and $0.5{\rm fm}$ at each temperature,
and examine the ground-state overlap by comparing $G(t)/G(0)$ with 
the fit function of $g_{\rm fit}(t)=A\cosh [m(t-N_t/2)]$ \cite{ISM02}. 
As a result, the optimal size seems to be $\rho \simeq 0.2{\rm fm}$ 
for $c \bar c$ systems. 
Hereafter, we only show the numerical results for $\rho=0.2{\rm fm}$. 

\section{Lattice QCD results for the $c\bar c$  system above $T_c$}
 
We investigate the $c\bar c$ systems above $T_c$ 
both in $J/\Psi (J^P=1^-)$ and $\eta_c (J^P=0^-)$ channels.
For each channel, 
we calculate the temporal correlator $G(t)$ and 
the effective mass $m_{\rm eff}(t)$ defined by Eq.(\ref{cosh}) both in PBC and APBC, 
and examine their spatial boundary-condition (b.c.) dependence.

Figures 1-4 show the effective-mass plot $m_{\rm eff}(t)$ 
of the $c\bar c$ system in the $J/\Psi$ channel for $\rho$=0.2fm.
From the cosh-type fit for the correlator $G(t)$ in the plateau region of $m_{\rm eff}(t)$, 
we extract the energies, $E$(PBC) and $E$(APBC), 
of the low-lying $c\bar c$ system in PBC and APBC, respectively.
Table 3 summarizes the $c\bar c$ system in the $J/\Psi$ channel in PBC and APBC 
at each temperature. 

As a remarkable fact, almost no spatial b.c. dependence is found 
for the low-lying energy of the $c \bar c$ system, 
i.e., $\Delta E \equiv E{\rm (APBC)}-E{\rm (PBC)} \simeq 0$, 
which is contrast to the $c \bar c$ scattering case of  
$\Delta E \simeq 2\sqrt{m_c^2+3\pi^2/L^2}-2m_c\simeq $ 0.35GeV 
for $L \simeq$1.55fm and $m_c \simeq$ 1.3GeV as was discussed in Sect.2.
This result indicates that $J/\Psi$ survives for $T=(1.11-2.07)T_c$.

Table 4 summarizes the $c \bar c$ system in the $\eta_c$ channel in PBC and APBC 
at each temperature.
Again, almost no spatial b.c. dependence is found as $\Delta E \equiv E{\rm (APBC)}-E{\rm (PBC)}\simeq 0$, 
and this result indicates that $\eta_c$ also survives for $T=(1.11-2.07)T_c$ 
as well as $J/\Psi$. 

In contrast to $J/\Psi$ and $\eta_c$, our preliminary lattice results show  
a large spatial b.c. dependence for the $c \bar c$ system 
in the $\chi_{c1}$ ($J^P=1^+$) channel even near $T_c$, 
which seems consistent with Ref.\cite{DKPW04}. 

\begin{figure}
\begin{center}
\begin{minipage}{7cm}
\rotatebox{-90}{\includegraphics[width=5cm]
{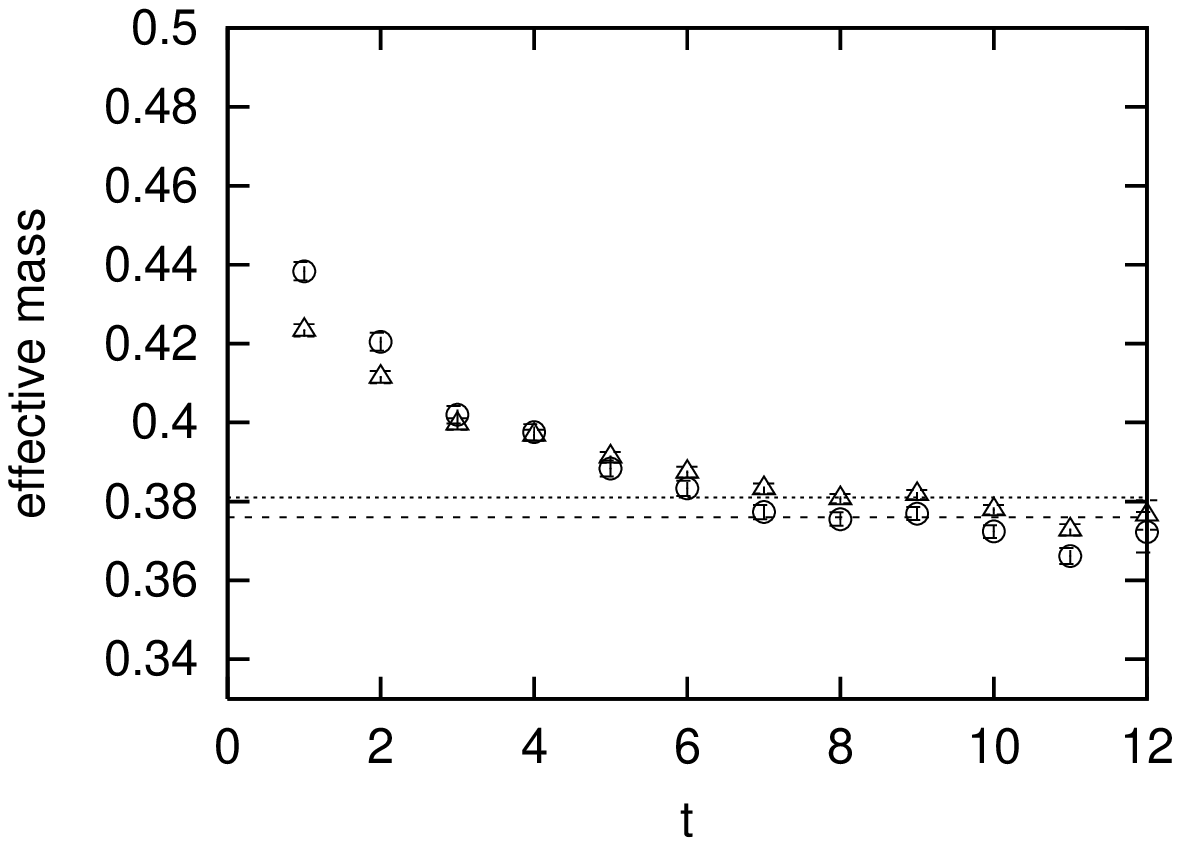}}
\vspace{-0.2cm}

\caption{Effective mass of $J/\Psi$ at $T=1.11T_c$ in lattice unit.
 The circular/triangle symbols denote the results in PBC/APBC.
 The dashed/dotted lines denote $E$(PBC/APBC) in PBC/APBC.
}
\label{fig1}
\end{minipage}
\begin{minipage}{7cm}
\rotatebox{-90}{\includegraphics[width=5cm]
{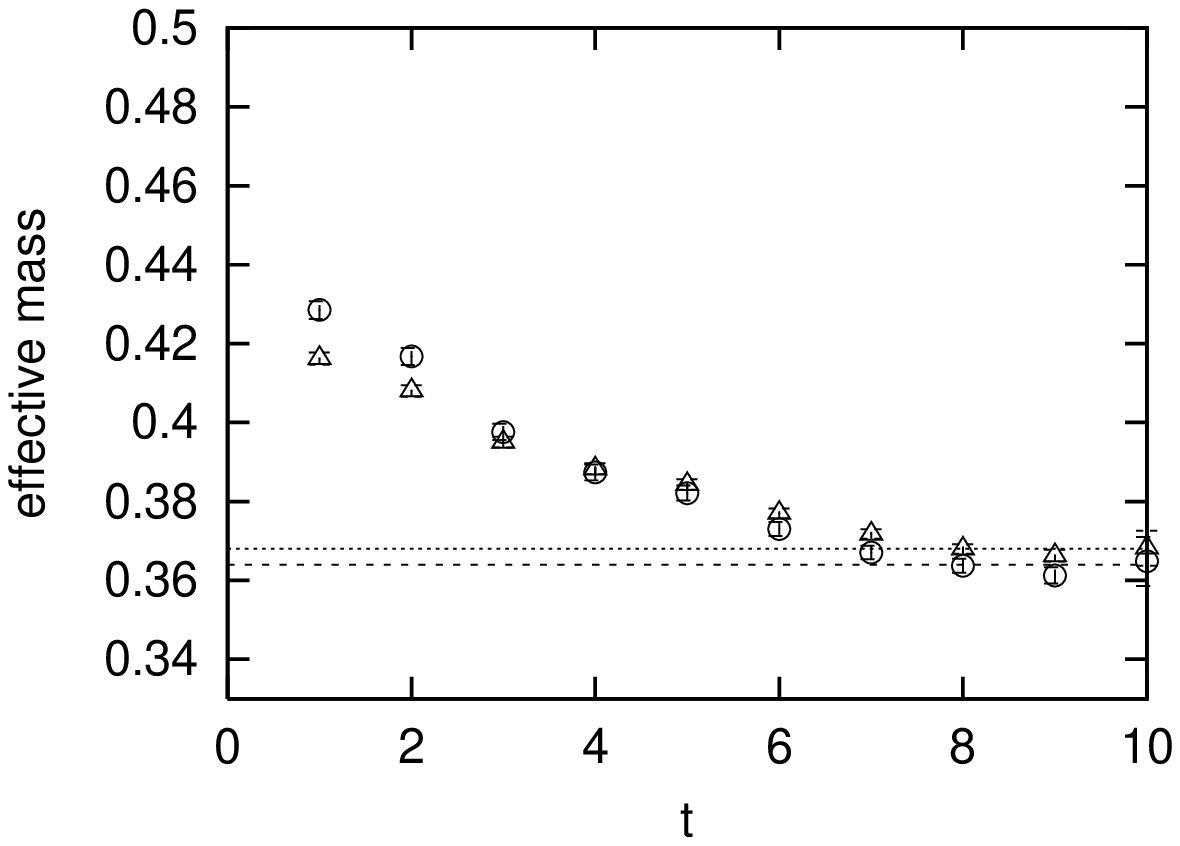}}
\vspace{-0.2cm}

\caption{Effective mass of $J/\Psi$ at $T=1.32T_c$. 
The symbols, lines, units are same as Figure 1.}
\label{fig2}
\end{minipage}
\end{center}
\end{figure}
\begin{figure}
\begin{center}
\begin{minipage}{7cm}
\rotatebox{-90}{\includegraphics[width=5cm]
{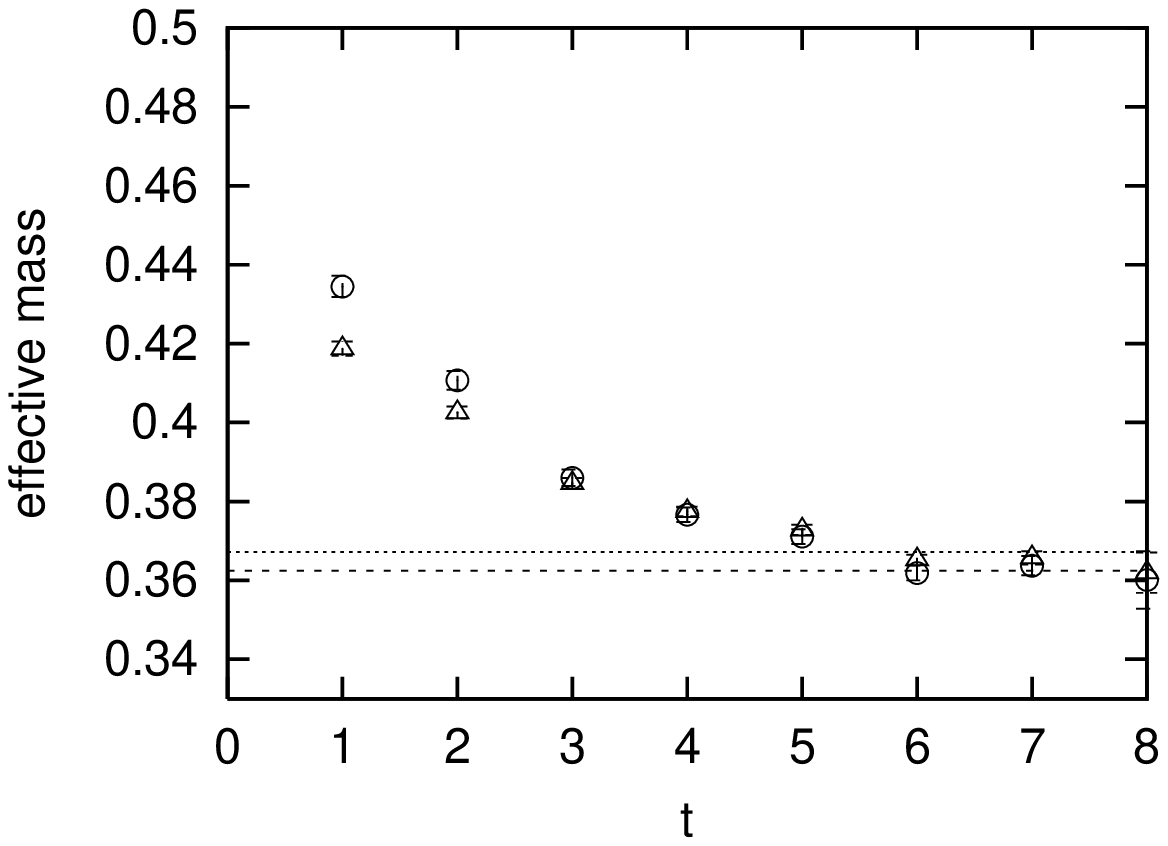}}
\vspace{-0.2cm}

\caption{Effective mass of $J/\Psi$ at $T=1.61T_c$. 
The symbols, lines, units are same as Figure 1.}
\label{fig3}
\end{minipage}
\begin{minipage}{7cm}
\rotatebox{-90}{\includegraphics[width=5cm]
{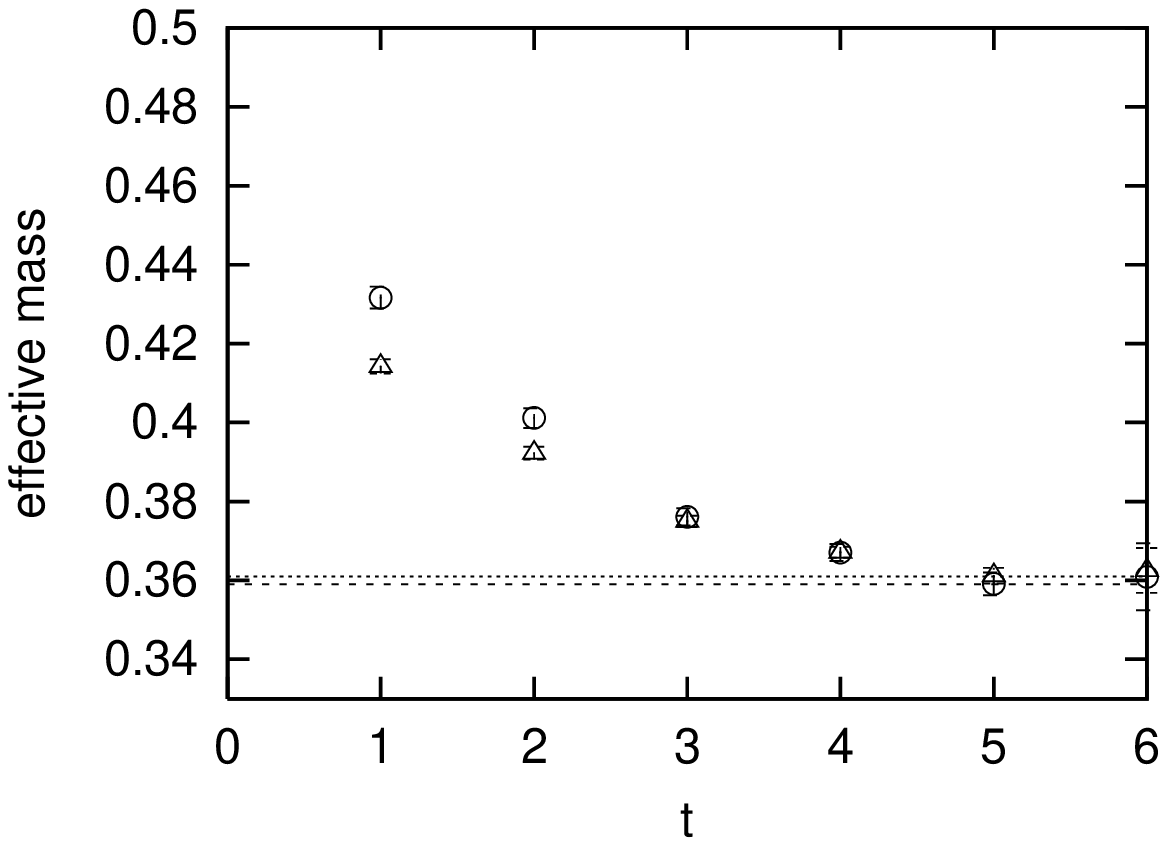}}
\vspace{-0.2cm}

\caption{Effective mass of $J/\Psi$ at $T=2.07T_c$. 
The symbols, lines, units are same as Figure 1.}
\label{fig4}
\end{minipage}
\end{center}
\end{figure}

\vspace{-0.18cm}
\section{Summary and conclusions}
We have investigated $J/\Psi$ and $\eta_c$ above $T_c$ with anisotropic 
quenched lattice QCD to clarify whether the $c\bar c$ systems above $T_c$ are 
compact quasi-bound states or scattering states.
We have adopted $O(a)$-improved Wilson quark action with renormalized anisotropy 
$a_s/a_t=4$. Anisotropic lattice is technically important for the measurement 
of temporal correlators at high temperatures.
We have use $\beta=6.10$ on $16^3\times (14-26)$ lattices, 
which correspond to $T=(1.11-2.07)T_c$. 

To conclude, we have found almost no spatial boundary-condition dependence of 
the energy of the low-lying $c\bar c$ system  
both in $J/\Psi$ and $\eta_c$ channels even on the finite-volume lattice. 
These results indicate that $J/\Psi$ and $\eta_c$ 
survive as compact $c \bar c$ quasi-bound states for $T=(1.11-2.07)T_c$.

\begin{table}
\begin{center}
\caption{The energy of the $c\bar c$ system in the $J/\Psi$ channel ($J^P=1^{-}$)
in PBC and APBC at $\beta=6.10$ and $\rho=0.2{\rm fm}$ at each temperature. 
The statistical errors are smaller than 0.01GeV.
We list also uncorrelated $\chi^2/N_{\rm DF}$ and 
$\Delta E\equiv E{\rm (APBC)}-E{\rm (PBC)}$.}
\vspace{0.2cm}
\label{data1}
\begin{tabular}{lllll}
\hline
\hline
temperature & fit range & $E$(PBC) [$\chi^2/N_{\rm DF}$] & $E$(APBC) [$\chi^2/N_{\rm DF}$]
 & $\Delta E$\\
\hline
$1.11T_c$ & 7--11  &3.05{\rm GeV} [0.14] & 3.09{\rm GeV} [0.61] &0.04{\rm GeV}
\\
$1.32T_c$ & 8--11 &2.95{\rm GeV} [0.34] & 2.98{\rm GeV} [0.33)] &0.03{\rm GeV}\\
$1.61T_c$ & 6--9  &2.94{\rm GeV} [0.10] & 2.98{\rm GeV} [0.22]
 &0.04{\rm GeV}\\
$2.07T_c$ & 5--7  &2.91{\rm GeV} [0.03] & 2.93{\rm GeV} [0.04]
 &0.02{\rm GeV}\\
\hline
\hline
\end{tabular}
\end{center}
\end{table}
\begin{table}
\caption{The energy of the $c\bar c$ system in the $\eta_c$ channel ($J^P=0^{-}$) 
in PBC and APBC at $\beta=6.10$ and $\rho=0.2{\rm fm}$ at each temperature.
 The statistical errors are smaller than 0.01GeV.
We list also uncorrelated $\chi^2/N_{\rm DF}$ and 
$\Delta E\equiv E{\rm (APBC)}-E{\rm (PBC)}$.}
\vspace{-0.2cm}
\label{data2}
\begin{center}
\begin{tabular}{lllll}
\hline
\hline
temperature & fit range & $E$(PBC) [$\chi^2/N_{\rm DF}$] & $E$(APBC)  
[$\chi^2/N_{\rm DF}$] & $\Delta E$\\
\hline
$1.11T_c$ &7--11  &3.03{\rm GeV} [0.04] & 3.02{\rm GeV} [0.17]  &-0.01{\rm GeV}\\
$1.32T_c$ & 7--11 &2.99{\rm GeV} [0.78] & 2.98{\rm GeV} [0.82] &-0.01{\rm GeV}\\
$1.61T_c$ & 6--9  &3.00{\rm GeV} [0.31] & 2.97{\rm GeV} [0.38] &-0.03{\rm GeV}\\
$2.07T_c$ & 5--7  &3.01{\rm GeV} [0.03] & 3.00{\rm GeV} [0.07] &-0.01{\rm GeV}\\
\hline
\hline
\end{tabular}
\end{center}
\end{table}

\end{document}